\documentclass[twocolumn,showpacs,prb,amsmath,amssymb,floatfix]{revtex4}
\usepackage{graphicx}
\usepackage{latexsym}
\usepackage{amsmath}
\usepackage{color}

\newcommand{\dis}{\displaystyle}
\newcommand{\nn}{\nonumber \\}
\newcommand{\bqu}{\begin{equation}}

\begin{document}

\title{%
Chirality Selection Models in a Closed System
}

\author{
Yukio \textsc{Saito}
 and Hiroyuki \textsc{Hyuga}}
\affiliation{\
Department of Physics, Keio University, Yokohama 223-8522, Japan 
}
\begin{abstract}
Recently the first chemical systems which show the amplification of
enantiomeric excess (ee) was found. Inspired by these experiments,
we propose a few chemical reaction models in a closed system.
The reactions consist of autocatalytic production of chiral molecules
and their decomposition processes.
By means of flow diagrams in the phase space of enatiometric concentrations,
chiral symmetry breaking is studied.

Without any destruction processes, the quadratic autocatalysis leads to 
the amplification of ee, but the flow is attracted to a fixed line,
which is only marginally stable.
The destruction processes alter the attractor structure completely.
With a cross inhibition the linear autocatalysis is sufficient for the
chiral symmetry breaking with a complete homochirality.
With a back reaction, the quadratic autocatalysis is necessary to bring about
the chiral symmetry breaking. The final ee is not complete
if there is a  racemization process.
\end{abstract}

\maketitle

\section{Introduction}
Because of
the three-dimensional tetragonal arrangement of
four chemical bonds around a carbon atom,
organic molecules can take two types of stereostructures; 
a right-handed one and a mirror-image left-handed one.
\cite{bonner88,feringa+99}
The molecules which have either of these two forms are called chiral, and
the two forms are enantiomers with each other.
In general the two types are classified in $R$ and $S$ forms, 
but, for sugars and 
amino acids, {\footnotesize D} and {\footnotesize L} representations are common in use.
From energetic point of view, the two enantiomers are indistinguishable and
they should exist with equal probability.
In fact, in the usual chemical reactions both types of 
enatiomers are produced with equal probabilities, and the product is 
called a racemate or a racemic mixture.

In contrast, life on earth utilizes only one type of enantiomers:
levorotatory({\footnotesize L})-amino acids  and 
dextrorotatory({\footnotesize D})-sugars.
This symmetry breaking in chirality is called the homochirality, and
the origin of this unique chirality selection has long been intriguing 
many scientists.
\cite{bonner88}
Various mechanisms have been proposed to induce an
asymmetry in the primordial molecular environment,
\cite{mason+85,kondepudi+85,meiring87,bada95,bailey+98,feringa+99,
hazen+01}
but the induced chiral asymmetry in all the proposed models 
is expected to be very minute.
Therefore, the mechanisms to amplify enormously the enatiometric excess (ee)
are indispensable. 

Frank is the  first to show that an autocatalytic 
chemical reaction 
with an antagonistic process in open systems
can lead to an amplification of ee.\cite{frank53}
Many theoretical models have since been proposed, 
but they are often criticized as lacking any experimental support.
\cite{bonner88}
Recently, a first system which shows the spontaneous amplification of  
ee has been found by Soai's group.
\cite{soai+90,soai+95,sato+01,sato+03}
This enhancement takes place in a closed system, and is 
explained by the second-order autocatalytic reaction.\cite{sato+03} 
Inspired by these experiments, \cite{soai+90,soai+95,sato+01,sato+03} we
constructed
some simple models of autocatalytic chemical reactions in a
closed and conserved system,
\cite{saito+04a,saito+04b} 
and studied
 conditions when the chiral symmetry breaking is possible.
Here we 
extend these models and study them
in detail.

In \S 2, we summarize the fundamental aspects of autocatalytic reaction
models where two 
enantiomers
 compete against each other 
in a closed system for the limited resource.\cite{saito+04a}
The ee amplification is found possible if 
the production of chiral molecules contains
the quadratic autocatalytic process.
However, 
concentrations of enantiomers are attracted to a fixed line,
which is only marginally stable. Therefore, the final configuration 
might be 
susceptible to external perturbation.
In \S 3, with the mutual antagonism or the cross inhibition,
the complete homochirality is concluded if there is a linear autocatalysis.
In \S 4, with the  back reaction, the quadratic autocatalysis 
is found necessary for the chirality selection.
Summary and discussions are given in the last section \S 5.

\section{Autocatalytic Reactions in a Closed System}
We start our study about reaction models with autocatalytic productions 
of chiral molecules from achiral materials.
Frank \cite{frank53} has considered the chiral symmetry breaking 
in an open chemical system.
However, since the experiment is performed in a closed system,
\cite{soai+95}
we consider 
reaction models in a closed and conserved system, and
explore effects that the autocatalysis alone 
can lead.

Suppose two achiral molecules A and B react to produce a chiral molecule C.
For the purpose of simplicity we denote ($R$)-C molecule as R
and ($S$)-C molecule as S, hereafter.
The reactions A+B$\rightarrow$ R and A+B$\rightarrow$ S proceed in a closed
system, and thus they stop when the materials A and/or B are exhausted.
Here we assume that there is an ample amount of material B, 
and its concentration can be regarded as
constant all through the reaction.
Then, the process depends only on the concentrations of A, R and S molecules,
which we denote $a,~r$ and $s$, respectively.
The conservation of the total number of participating molecules 
gives the relation that the total concentration 
stays constant determined by the initial values as
\begin{align}
a+r+s=a_0+r_0+s_0=c_0=\mbox{const.}
\end{align}
Then the reaction developement can be depicted 
as a flow in the $r$-$s$ phase diagram.
The conservation law (1) limits the phase space in the triangular region
\begin{align}
0 \le r,s \le c_0, \quad 0 \le r+s \le c_0 .
\label{eq2}
\end{align}

As a first step we consider the simplest autocatalytic reaction model
such that the production from 
the material A to R or S is described by the following rate equations:
\begin{align}
\frac{d r}{dt} = f(r)a=f(r)(c_0-r-s)
\nonumber  \\
\frac{d s}{dt} = f(s)a=f(s)(c_0-r-s)
\label{eq3}
\end{align}
with 
\begin{align}
f(x)=k_0+k_1x+k_2x^2 .
\end{align}
The coefficient $k_0$ represents the rate of a nonautocatalitic reaction,
$k_1$ of a linearly autocatalytic 
and $k_2$ of a quadratically autocatalytic reactions.
Since all the rates $k_{0,1,2}$ are non-negative,
$f(x) \ge 0$ for $x=r$ and $s$.
In the Soai reaction, the quadratic autocatalysis is shown to explain the
time evolution of the reaction yield and the ee.\cite{sato+03}

\subsection{Fixed points}

We now study the flow diagram in $r-s$ space in typical situations.
First we search fixed points of Eq.(\ref{eq3}) in the region (\ref{eq2}).
It is easy to find that $a=0$ gives a line of fixed points at $r+s=c_0$, 
which are stable 
and correspond to the state where all the material A is consummed up.
Only for the case without nonautocatalysis, $k_0=0$,
another fixed point is possible at the origin $r=s=0$, but it is
easily shown to be unstable.

\begin{figure}[h]
\begin{center} 
\includegraphics[width=0.32\linewidth]{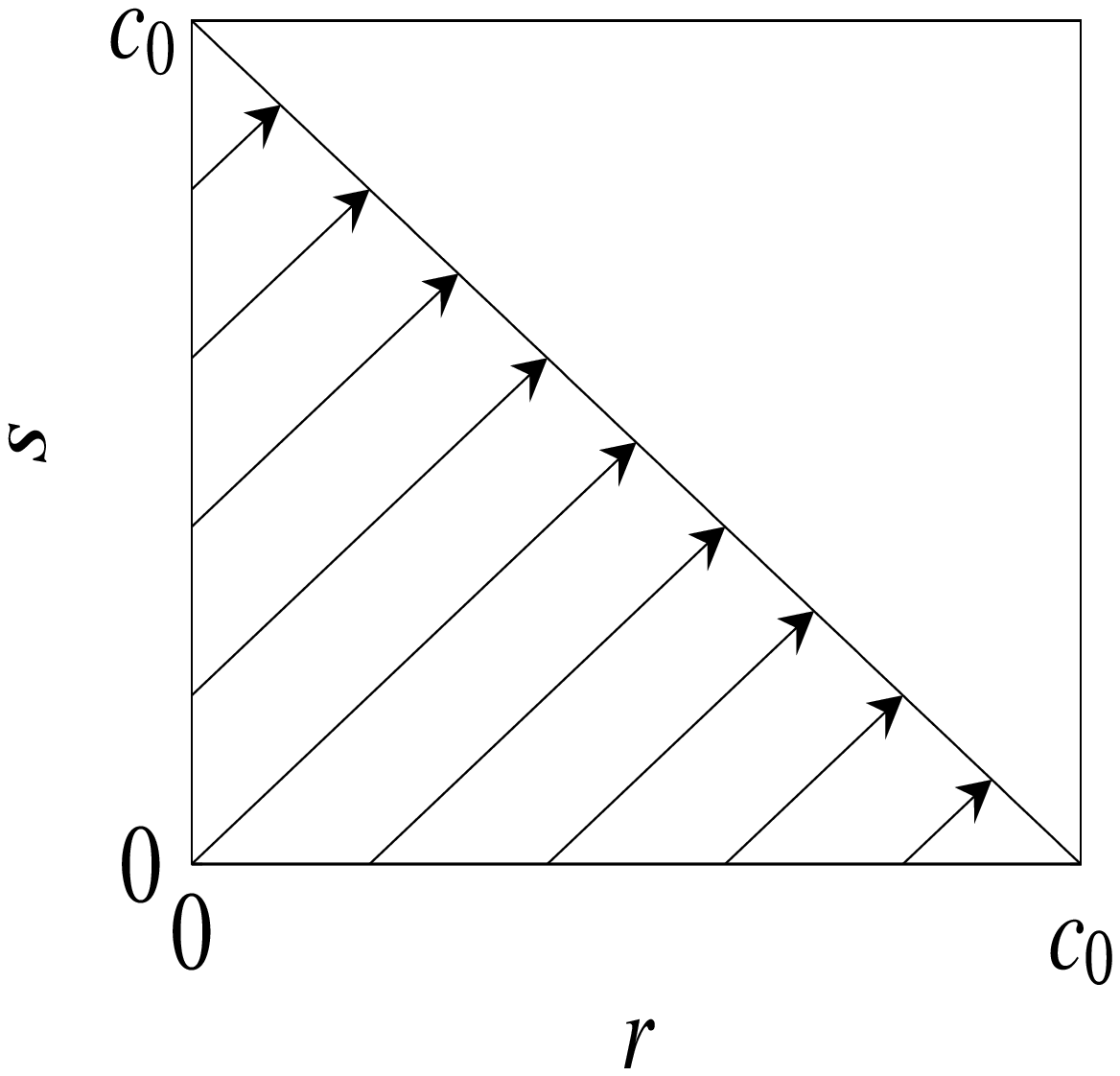}
\includegraphics[width=0.32\linewidth]{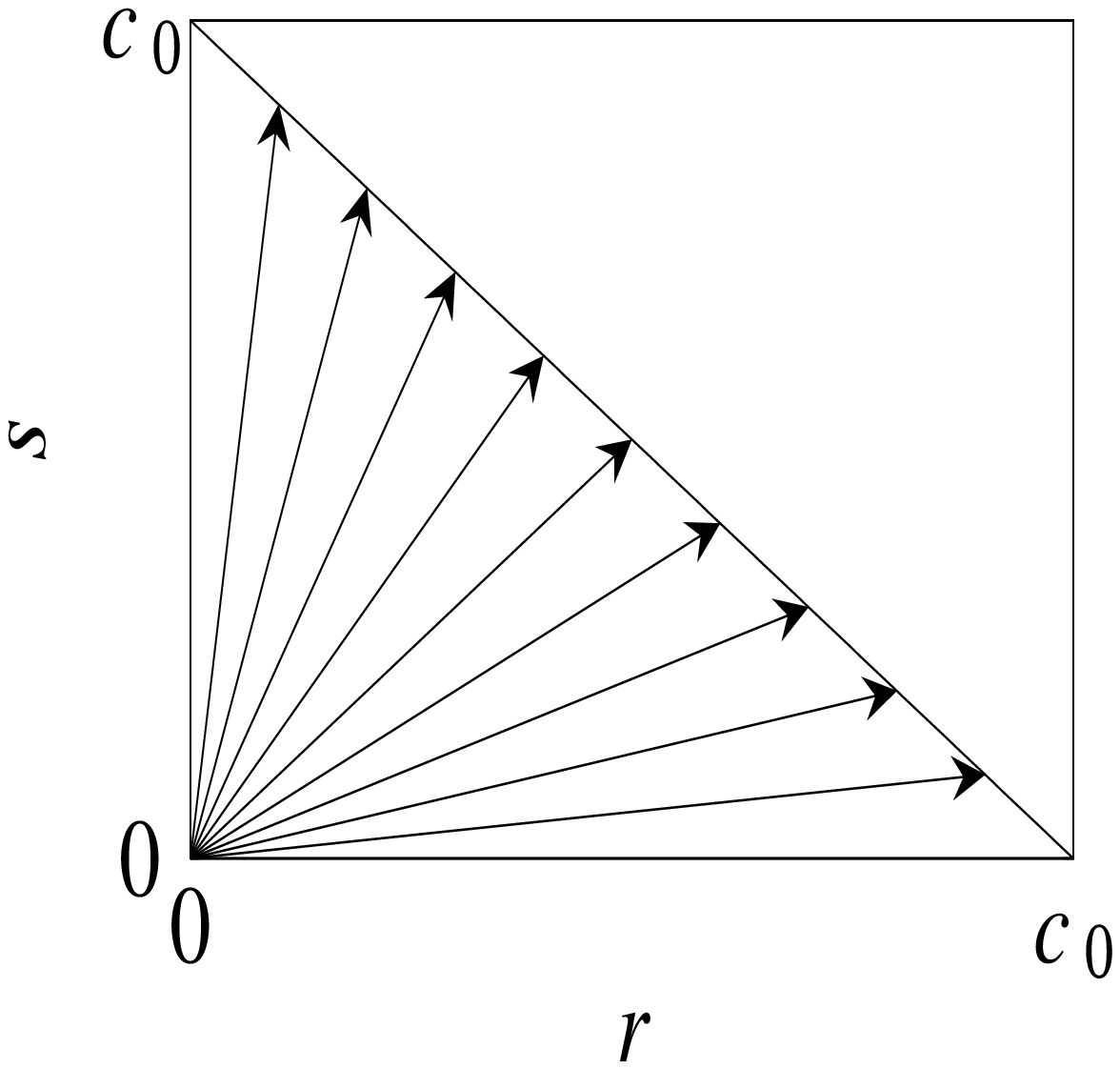}
\includegraphics[width=0.32\linewidth]{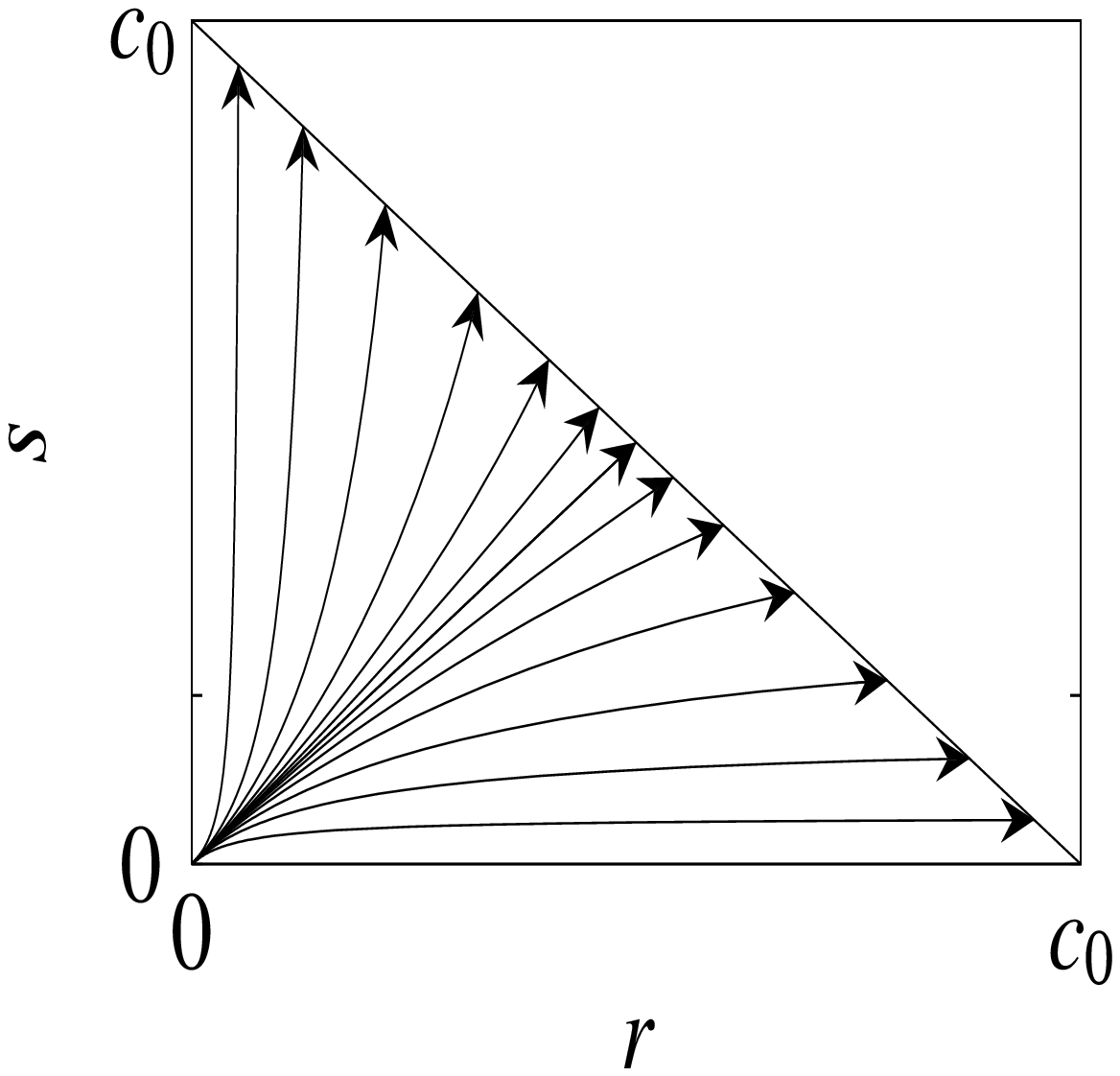}
\end{center} 
\caption{Flow diagrams for (a) nonautocatalytic ($k_0 >0,~k_1=k_2=0$), 
(b) linearly autocatalytic ($k_1>0,~k_0=k_2=0$),
and (c) quadratically autocatalytic ($k_2>0,~k_0=k_1=0$) reactions.
}
\label{fig1}
\end{figure}

The flow will then be attracted to the fixed line $r+s=c_0$, and is given by the
equation
\begin{align}
\int \frac{d r}{f(r)} - \int \frac{d s}{f(s)} = \mathrm{constant} .
\end{align}
Integration of this equation for general $f(x)$ can be carried out analytically,
but we discuss the flow lines only in three typical cases in the following.\\
$\bullet$\,{\bf Nonautocatalytic case:}

Without autocatalysis, $k_0 > 0$ but $k_1=k_2=0$, flow lines 
are easily obtained by interation of eq.(5) and they have
 a unit slope, $r-s=r_0-s_0=$constant, as shown in Fig.\ref{fig1}(a).\\ 
$\bullet$\,{\bf Linearly autocatalitic case:}

With only a linear autocatalysis, $k_1>0,~k_0=k_2=0$, 
the integration of eq.(5) is easy to yield 
the trajectory $r/s=r_0/s_0=$constant.
The flow is represented by lines projecting out from the origin, 
as shown in Fig.\ref{fig1}(b).\\
$\bullet$\,{\bf Quadratically autocatalytic case:}

For a quadratically autocatalytic case, $k_2 >0, ~k_0=k_1=0$,
flow lines are a collection of hyperbolae projecting out from the origin,
$1/r-1/s=1/r_0-1/s_0=$constant, as shown in Fig.\ref{fig1}(c).

\subsection{Enatiometric Excess}

In order to quantify the ee or the degree of chiral symmetry breaking, 
we use an order parameter $\phi$ defined as
\begin{align}
\phi= \frac{r-s}{r+s} .
\end{align}
$\bullet$\,{\bf Nonautocatalytic case:}

For the non-autocatalytic case in Fig.1(a), 
the numerator $r-s$ remains constant during the time evolution whereas 
the denominator $r+s$ increases. Thus the ee decreases monotonously from
the initial value $|\phi_0|=|(r_0-s_0)/(r_0+s_0)|$
to the asymptotic one $|\phi_{\infty}|=|(r_0-s_0)/c_0|$.
If the initial concentrations $r_0$ and $s_0$ are small, 
ee $|\phi|$ becomes very small.
This result is quite natural from the microscopic point of view,
since the nonautocatalytic reaction
produces both enatiomers randomly and leads to the racemization.
\\
$\bullet$\,{\bf Linearly autocatalytic case:}

For the linearly autocatalytic case in Fig.1(b),
the ratio $r/s$ remains constant and so does the ee. \\
$\bullet$\,{\bf Quadratically autocatalytic case:}

Only with a nonlinear autocatalysis in Fig.1(c), 
the ee increases up to some saturation value $|\phi_{\infty}|$, as shown in Fig.2. 
This asymptotic value of $\phi_{\infty}$ depends on
the initial concentrations $r_0$ and $s_0$,
even when the initial $\phi_0$ and the total $c_0$ are fixed, 
as shown in Fig.2. 
The smaller the initial amount of chiral molecules  ($r_0+s_0$) are,
the more the ee is amplified.

\begin{figure}[h]
\begin{center} 
\includegraphics[width=0.6\linewidth]{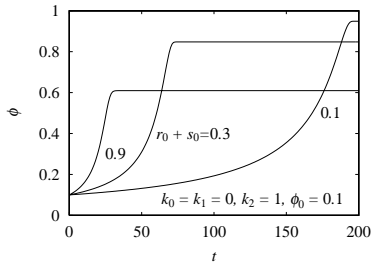}
\end{center} 
\caption{Time evolution of the enantiomeric excess $\phi$ 
with quadratic autocatalytic reaction $k_2>0$ with the same initial
value $\phi_0$ but the different amount of initial R and S molecules,
$r_0$ and $s_0$. 
}
\label{fig2}
\end{figure}

\noindent
$\bullet$\,{\bf General case:}

For more general cases,  the ee amplification is explained by studying 
the evolution of $\phi$, which is derived from eq.(3) as 
\begin{align}
\frac{d \phi}{dt} = \frac{a}{c_0-a} \phi 
[-2k_0 + \frac{k_2}{2} (c_0-a)^2 (1- \phi^2)] .
\label{eq7}
\end{align}
The evolution equation turns out to be independent of the
linear autocatalytic rate $k_1$.
Without quadratic autocatalysis ($k_2=0$) the ee decreases 
as we have already observed,
since
the concentration of A material, $a$, and the sum of R and S products,
$c_0-a=r+s$, should always be positive.
The situation does not alter by including only a linear autocatalysis with 
$k_1 >0$:
it does not directly affect the evolution of $\phi$.
Only when the quadratic auatocatalysis dominates at $k_2 > 4k_0/c_0^2$,
the ee can increase. 
Therefore, in this simplest model only the quadratic autocatalysis 
can 
enhance the chiral assymmetry.
This may correspond to the situation observed in the Soai reaction.
It has to be noted that the asymptotic value $\phi_{\infty}$ 
cannot be determined from eq.(7),
since 
$\dis\lim_{t \rightarrow \infty} a(t)=0$ 
in the present model.

Even though 
the amplification of the enatiometric excess or the amplification of 
the chiral assymmetry can be brought about
 by the nonlinear autocatalysis,
the final state of the system is on the line of fixed points,
and thus the final values of ee can be anything depending on 
the initial condition.
Since these points are marginally stable, i.e. they have no quaranteed 
stability 
along the direction of the fixed line, they may be susceptible to additional
perturbations.
We now consider the effect of two types of destructive reactions 
of chiral molecules.

\section{With Cross Inhibition}
Frank has introduced the mutual antagonism between the enantiomers, R and S,
and succeeded in explaining the chiral symmetry breaking in an open reaction
system.
Here we extend his model to the closed system. 
The reaction equation now incorporates terms of cross inhibition as
\begin{align}
\frac{d r}{dt} = f(r)a- \mu r s
\nonumber  \\
\frac{d s}{dt} = f(s)a - \mu r s .
\label{eq8}
\end{align}

\subsection{Fixed points}
Fixed points have to satisfy the relation
\begin{align}
f(r)a=f(s)a=\mu rs.
\end{align}
The simplest solution is $a=0$ and $rs=0$, which corresponds
to the end corners of the triangular region, $(r,~s)=(c_0,~0)$
and $(0,~c_0)$.  These are easily shown to be  stable with complete
homochirality, $\phi=\pm 1$.
Other fixed points are also possible 
with a nonzero $a$ and $f(r)=f(s)=\mu rs/a$. 
We shall study them in various cases below.
\\
$\bullet$\;{\bf Nonautocatalytic case:}

Only with a nonautocatalytic reaction ($k_0>0,~k_1=k_2=0$), 
$f(r)=f(s)=k_0$ and 
there is a fixed line
\begin{align}
\dis{(r+\frac{k_0}{\mu})(s+\frac{k_0}{\mu})
=\frac{k_0}{\mu}(c_0+\frac{k_0}{\mu})}
\label{eq10}
\end{align}
 passing through the two terminal points $(r,~s)=(c_0,~0)$
and $(0,~c_0)$.
This line is an accumulation of infinitely many marginal fixed points.
In fact the flow trajectory is easily integrated as $r-s=r_0-s_0=$constant,
the same lines with a unit slope obtained in the
previous section without cross inhibition. 
The cross inhibition only shifts the fixed line 
from the border $r+s=c_0$ in Fig.1(a) to the hyperbola (\ref{eq10}).

\begin{figure}[h]
\begin{center} 
\includegraphics[width=0.33\linewidth]{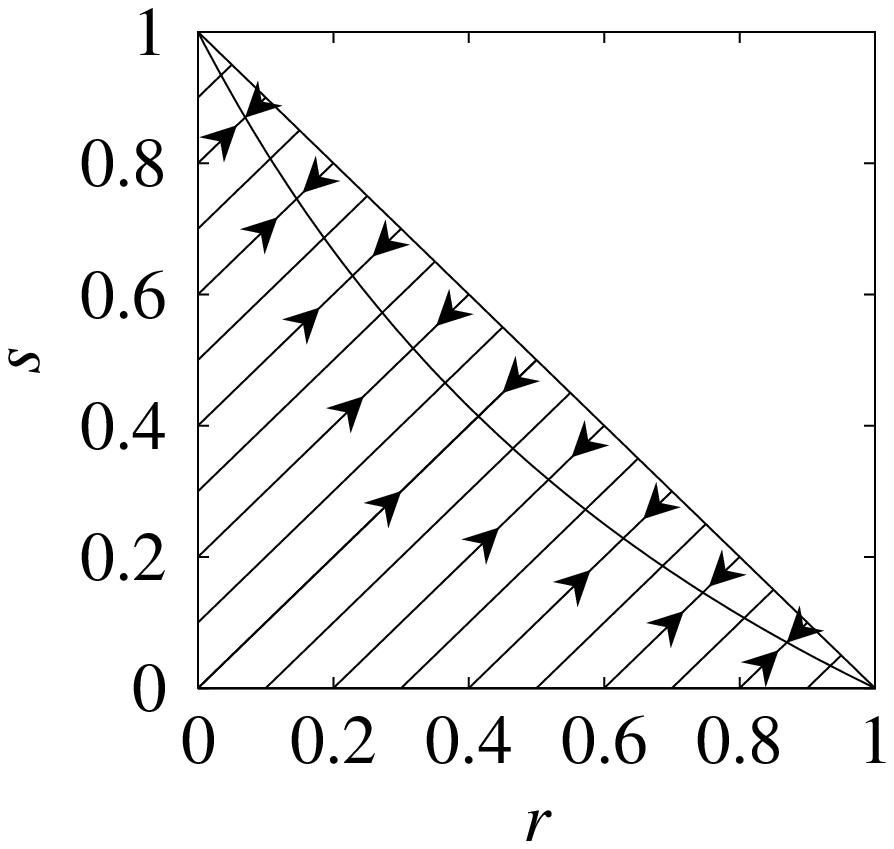}
\includegraphics[width=0.33\linewidth]{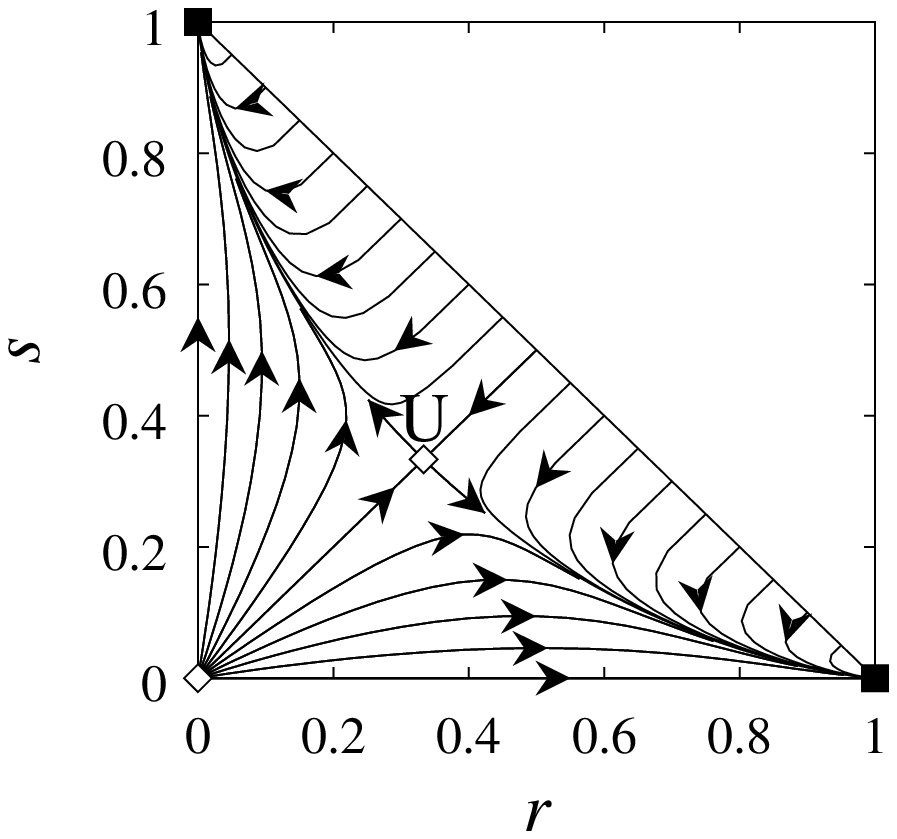}
\end{center} 
\caption{Flow diagrams with cross inhibition $\mu c_0=1$
at (a) $k_0 >0,~k_1=k_2=0$ and (b) $k_1>0,~k_0=k_2=0$.
}
\label{fig3}
\end{figure}

\noindent
$\bullet$\;{\bf Linearly autocatalytic case:}

With a finite $k_1$ as well as $k_0$, a fixed line is replaced by
a symmetric fixed point U at 
\begin{align}
r=s=
\frac{-2 k_0 +k_1c_0 + \sqrt{(2k_0+k_1 c_0)^2+4k_0c_0 \mu}}{2(\mu + 2k_1)} .
\label{eq11}
\end{align}
At $k_0=0$, there appears an additional fixed point at the origin $r=s=0$.
All these fixed points are easily shown to be unstable, as shown in Fig.3(b).
The only stable fixed points are those at the corners, $(c_0,0)$ and $(0,c_0)$,
with
the complete homochirality. 
Thus, with only a linear autocatalysis and without any nonlinear ones,
complete homochirality is possible with a cross inhibition 
even in a closed system.

With quadratic autocatalysis, the overall structure of 
the flow is similar to the case  with a linear autocatalysis.

\subsection{Enantiomeric excess}

Up to the linear autocatalysis ($k_2=0$) with a cross inhibition ($\mu \ne 0$),
the ee follows the evolution equation
\begin{align}
\frac{d \phi}{dt} = \frac{1}{c_0-a} \phi 
[-2k_0 a + \frac{\mu}{2} (c_0-a)^2 (1- \phi^2)] .
\label{eq12}
\end{align}
The rate $k_1$ of the linear autocatalytic process is not involved in the
symmetry breaking effect of $\phi$ neither in the present case
with cross inhibition 
nor in the case without cross inhibition in the previous section.
One notices the similarity between the evolution equation (\ref{eq12}) and
eq.(\ref{eq7}) derived in the previous section.
The effect of the quadratic autocatalytic reaction, $k_2$, 
in eq.(\ref{eq7}) is replaced by $\mu/a$ in eq.(\ref{eq12}).
In the present case, the symmetry recovering effect $k_0$ weakens as the
material A is consumed ($a \rightarrow 0$), 
and the complete homochirality $\phi= \pm 1$ is achieved asymptotically.
For example, the time evolution of $r/c_0,~s/c_0$ and $\phi$ 
is depicted in Fig.4
at $k_0=0, k_1c_0=1, \mu c_0=1$ and 
$r_0/c_0=0.0714, s_0/c_0=0.0700, \phi_0=0.01$.
Initially $r$ and $s$ increase simultaneously to approach 
the symmetric fixed point U: $(r/c_0, ~s/c_0)$=(1/3,1/3) 
with a constant $\phi \approx \phi_0$, 
driven by the linear autocatalytic process $k_1$.
However, since U is unstable, the trajectory is eventually expelled from U and 
$r$ increases at the cost of $s$. As a result $\phi$ increases to the full
homochirality.

\begin{figure}[h]
\begin{center} 
\includegraphics[width=0.60\linewidth]{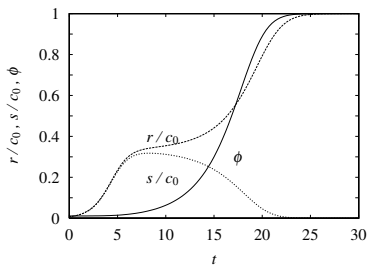}
\end{center} 
\caption{Time evolution of $r, s$, and $\phi$ 
at $k_0=0, k_1c_0=1, \mu c_0=1$ and $r_0/c_0=0.0714, s_0/c_0=0.0700, \phi_0=0.01$.
}
\label{fig4}
\end{figure}

\section{With Back Reactions}

We now consider another type of decomposition process, a simple back reaction
from the product R or S to an A molecule.
Governing equations are 
\begin{align}
 \dot{r}&=f(r)a-\lambda r\nn
 \dot{s}&=f(s)a-\lambda s
\end{align}
Since the back reaction is linear in the product concentrations $r$ and $s$, 
the evolution of the ee remains the same with eq.(7).
Therefore, the quadratic autocatalysis with a finite $k_2$ is the necessity
to break the chiral symmetry in the present case.
Even though $\phi$ follows the same evolution (7) without
the back reaction, its asmptotic behavior is quite different from that
in \S 2, because the fixed point structure is drastically altered by the
back reaction.

\noindent\large
{\bf Fixed points}\normalsize  

Fixed points of the flow satisfy conditions
\bqu 
f(r)a-\lambda r=f(s)a-\lambda s=0.
\end{equation}
It is trivial that $a=0$ cannot be a fixed point 
any more unless $c_0=0$.
As we have supposed, 
the fixed line $r+s=c_0$ in \S 2 
is very sensible to the perturbation like a back reaction.
We now study structures of fixed points in some simple cases.

\noindent
$\bullet$\,{\bf Nonautocatalytic case:}

In the non-autocatalytic case with $k_0>0,~k_1=k_2=0$ with a finite $\lambda$,
there is only a symmetric fixed point at 
\begin{align}
r=s=\frac{c_0/2}{1+\lambda/2k_0} ~<~\frac{c_0}{2},
\end{align} 
which is easily shown to be stable, as shown in Fig.5(a).

\begin{figure}[h]
\begin{center} 
\includegraphics[width=0.32\linewidth]{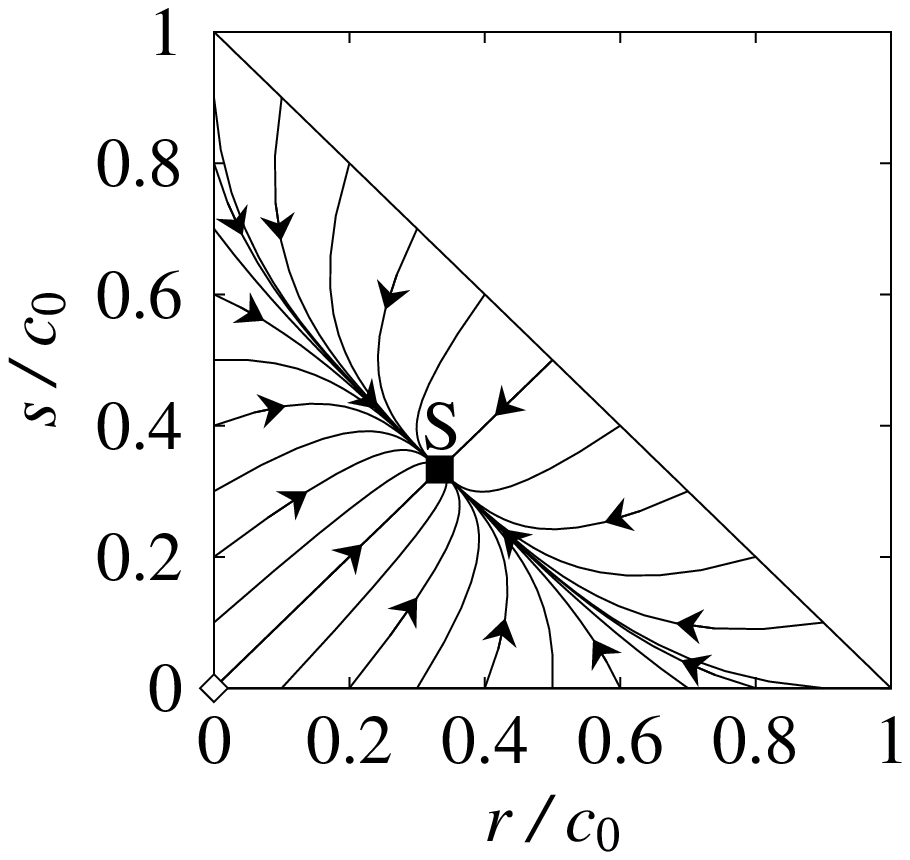}
\includegraphics[width=0.32\linewidth]{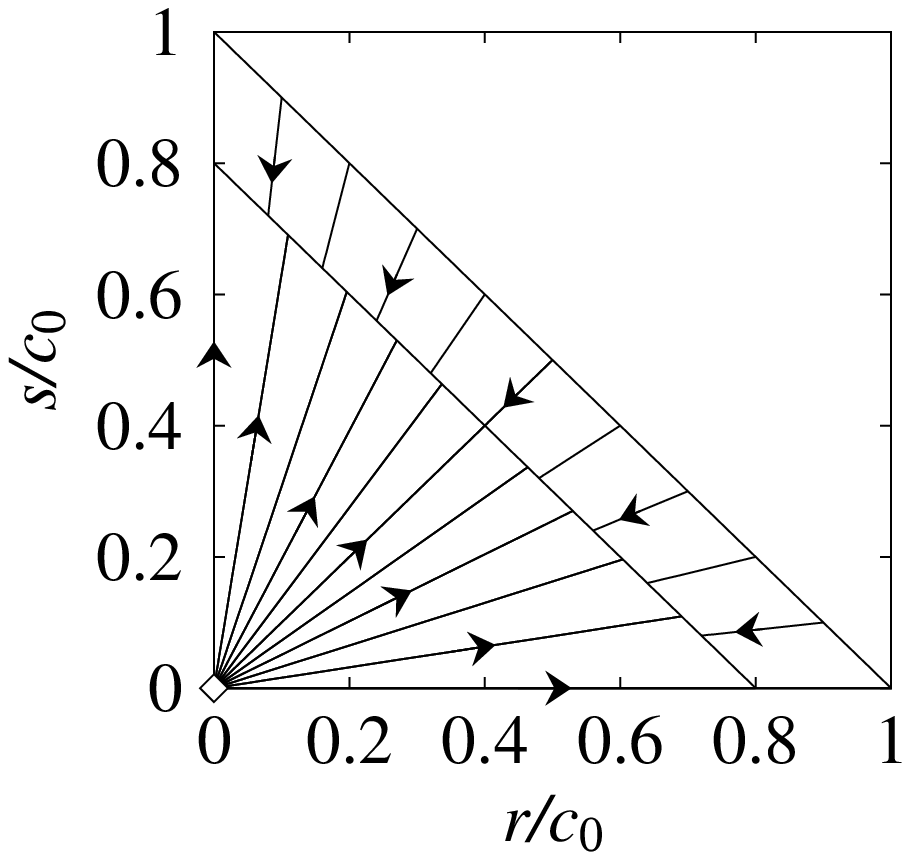}
\includegraphics[width=0.32\linewidth]{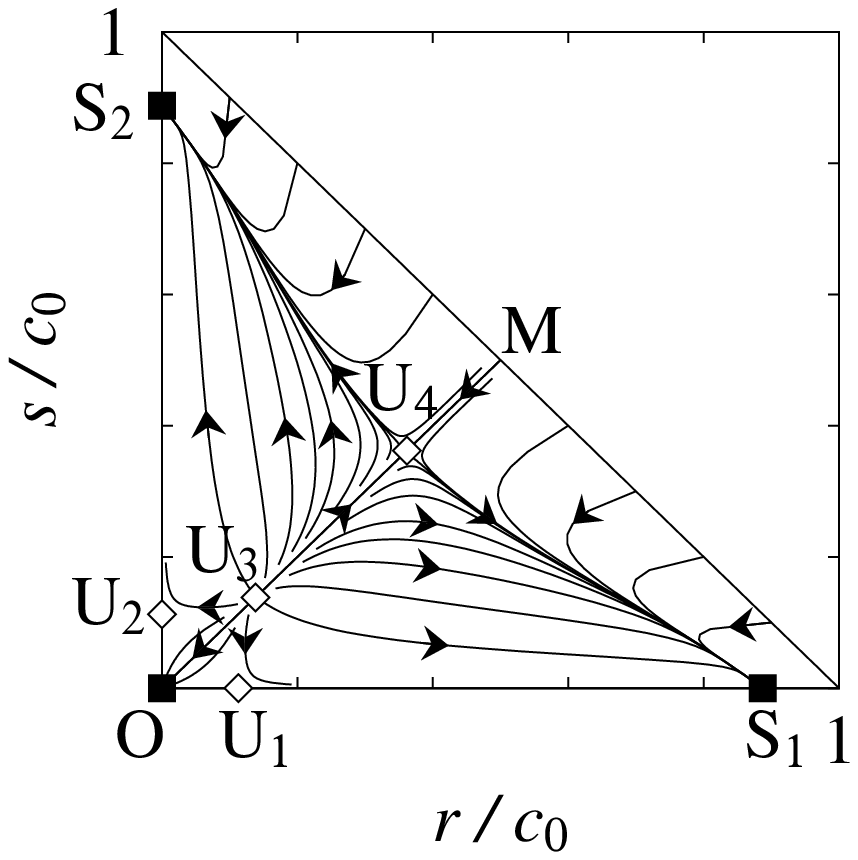}
\includegraphics[width=0.32\linewidth]{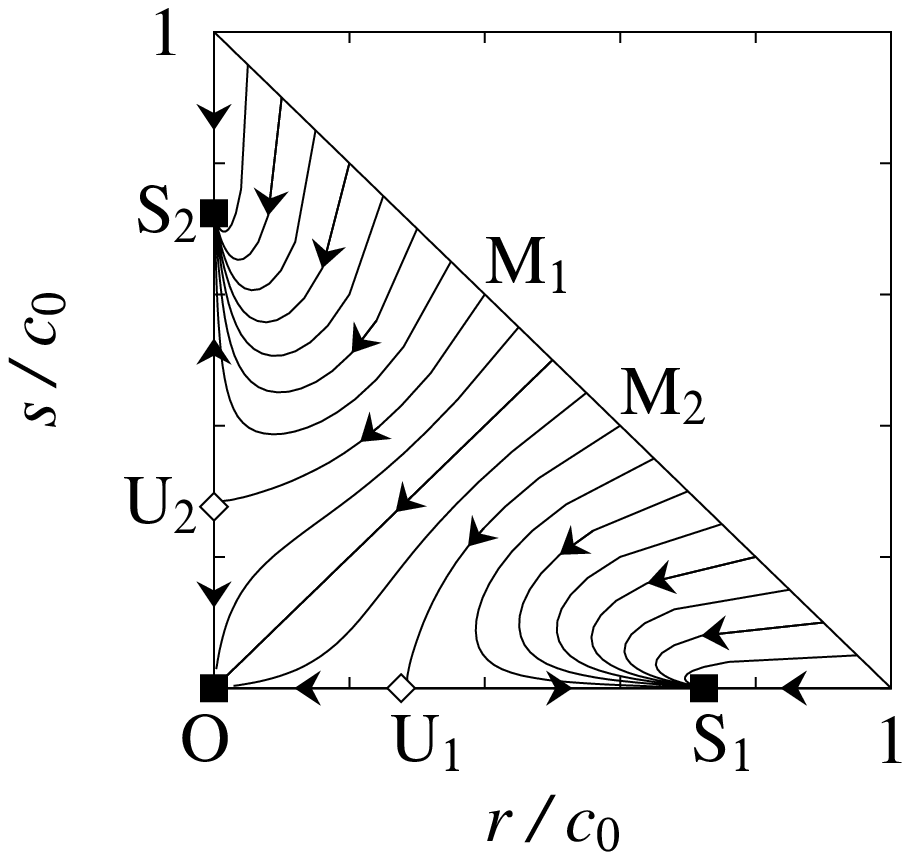}
\includegraphics[width=0.32\linewidth]{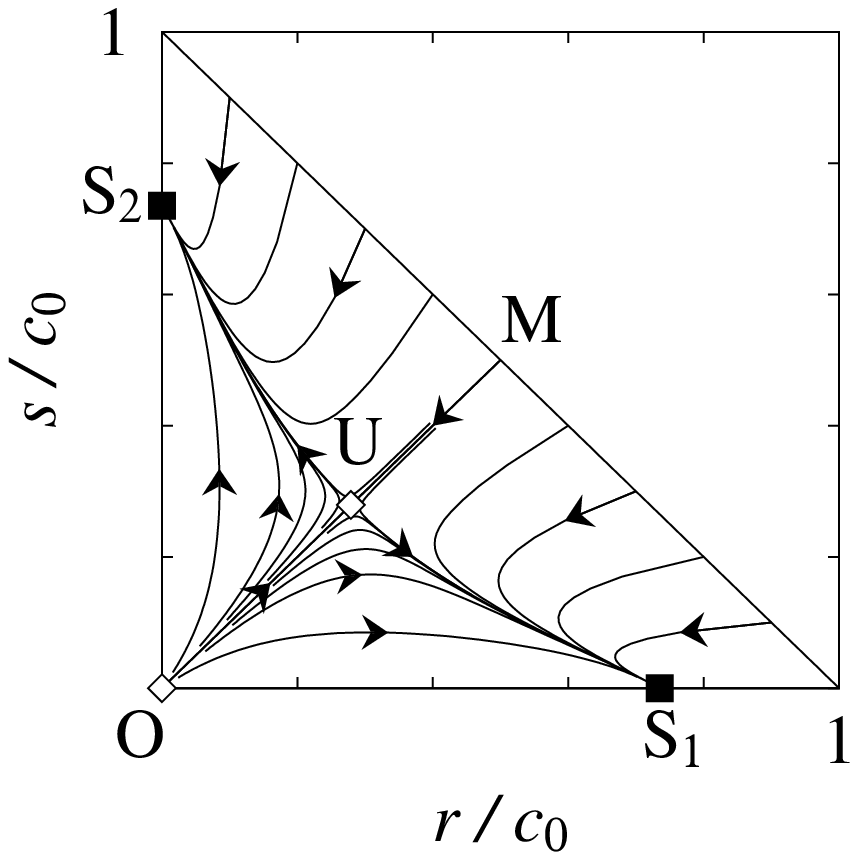}
\includegraphics[width=0.32\linewidth]{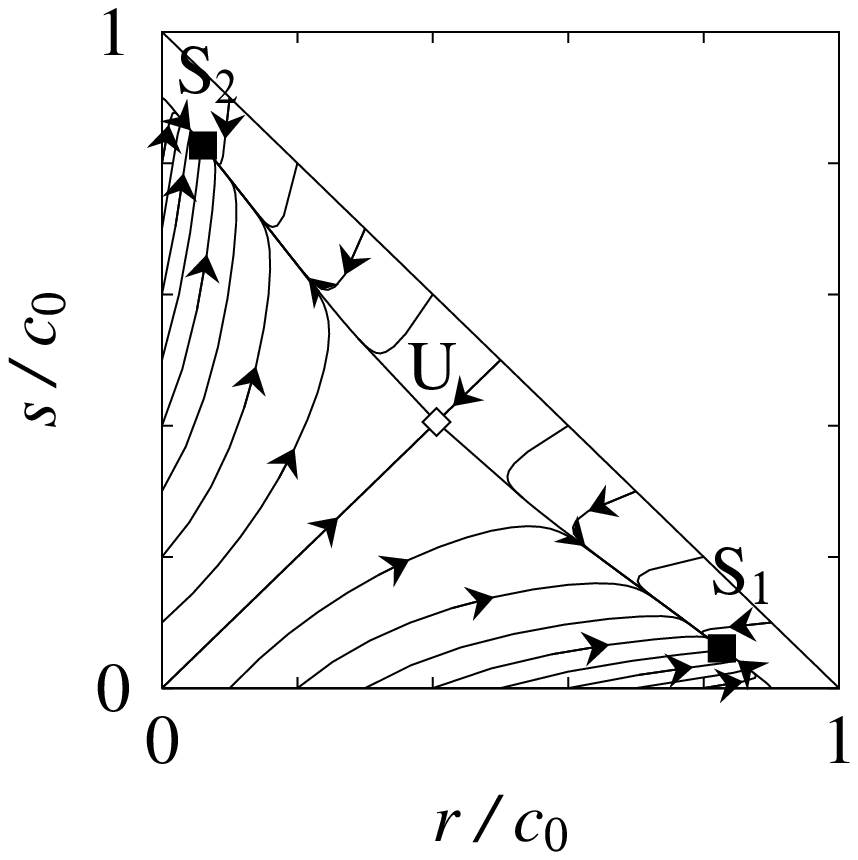}
\end{center} 
\caption{Flow diagrams for 
(a) $k_0 >0,~k_1=k_2=0, \lambda=1$, (b) $k_1=1, k_0=k_2=0, \lambda=0.1$,
(c) $k_2 =1,~k_0=k_1=0, \lambda=0.1$, (d) $k_2=1,~k_0=k_2=0, \lambda=0.2$,
(e) $k_1c_0=0.4, k_2 c_0^2 =1,~k_0=0, \lambda=0.3$.
(f) $k_0=0.05, k_1c_0=0., k_2 c_0^2 =1, \lambda=0.1 $.
}
\label{fig5}
\end{figure}

\begin{figure}[h]
\begin{center} 
\includegraphics[width=0.45\linewidth]{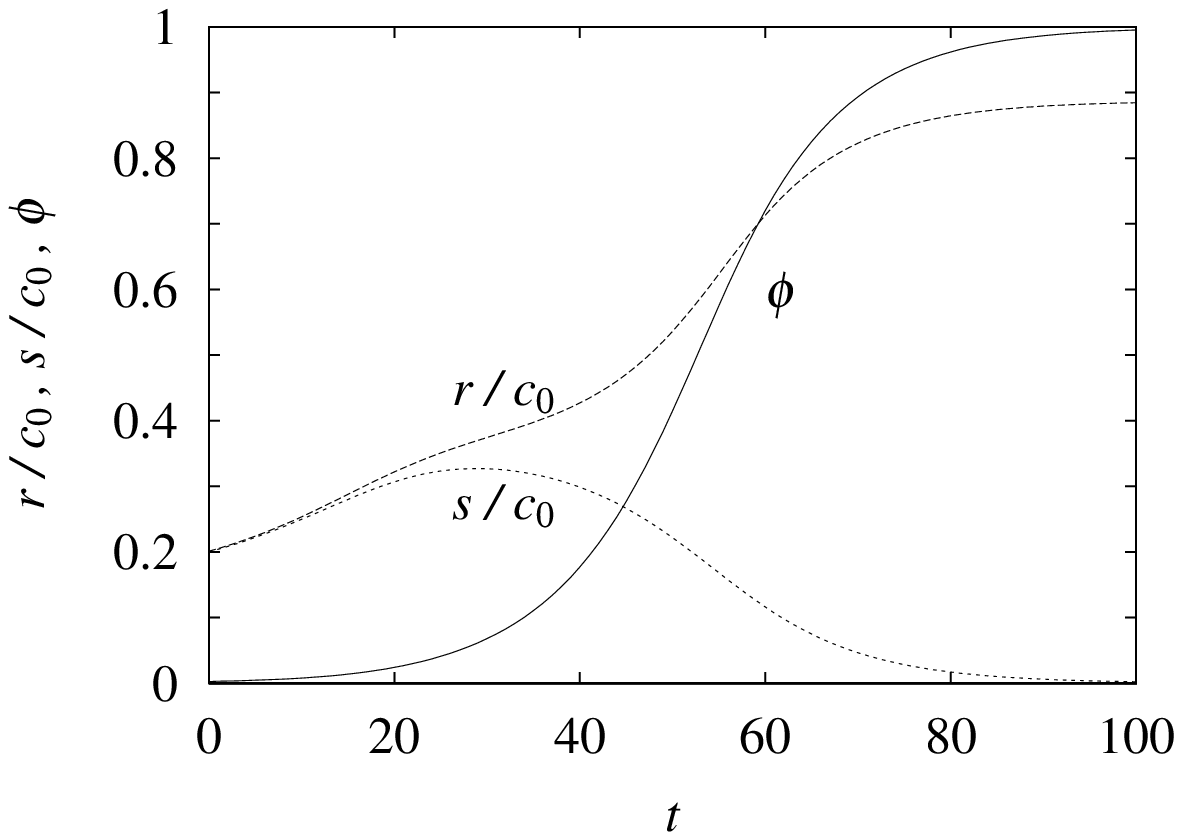}
\includegraphics[width=0.45\linewidth]{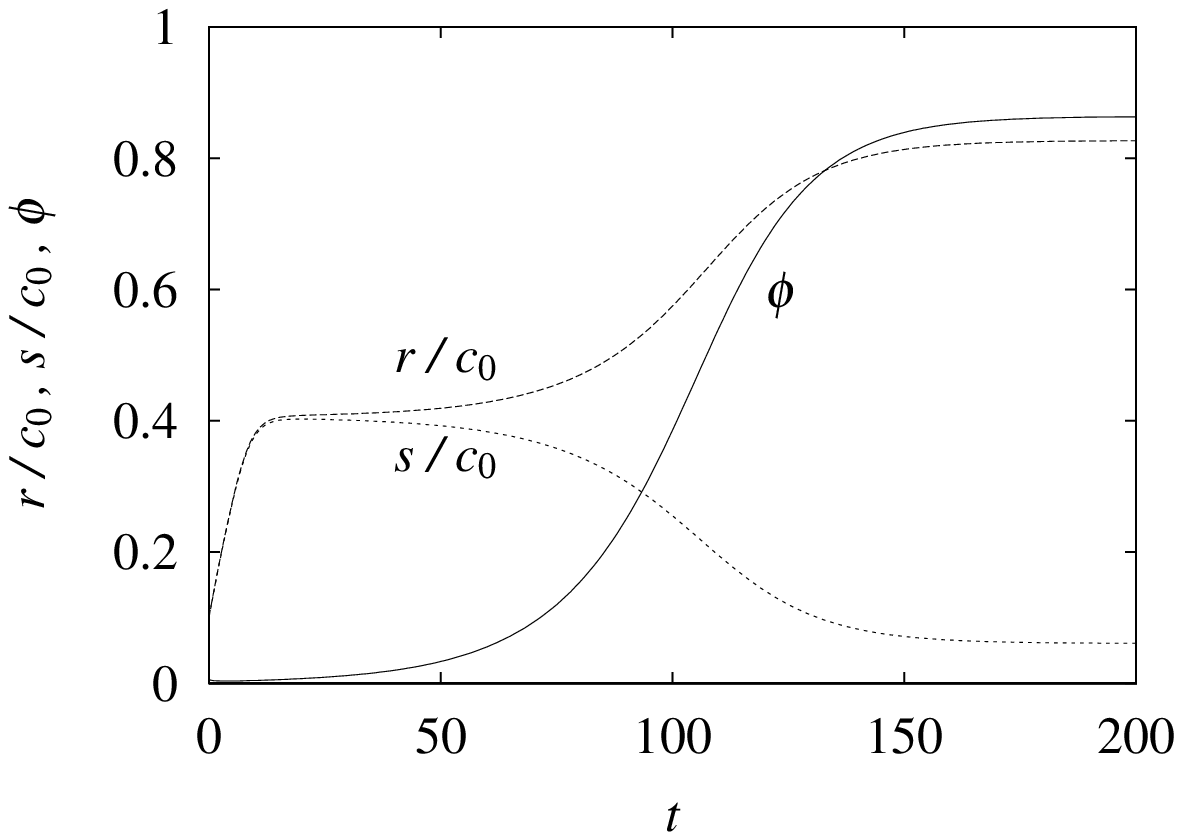}
\end{center} 
\caption{Time evolution of the ee $\phi$ with nonlinear autocatalysis and
a back reaction (a) without random creation $k_0=0$, and (b) with
a finte $k_0=0.05$.
}
\label{fig6}
\end{figure}

\noindent
$\bullet$\,{\bf Linearly autocatalytic case:}

With only a linear autocatalysis $k_1>0, ~k_0=k_2=0$,
we have only a trivial fixed point at the origin $(r,s)=(0, 0)$,
when the rate of back reaction is as large as $\lambda \ge k_1 c_0$: 
there will be no net production of chiral C molecules 
because of strong back reaction.
On the contrary, 
when the rate of back reaction is small as $\lambda < k_1c_0$, 
there are additional fixed points on a line $r+s=c_0-\lambda/k_1$.
This line corresponds to the one $r+s=c_0$ in \S 2 
shifted by the back reaction $\lambda$. 
In fact the trajectory 
is the same as in \S 2 and
radiates from the origin $r/s=r_0/s_0=$constant, as shown in Fig.5(b).

\noindent
$\bullet$\,{\bf Quadratically autocatalytic case:}

With a quadratic autocatalysis $k_2 >0, k_0=k_1=0$, the back reaction
decimates the fixed line $r+s=c_0$ and produces several fixed points.
One of them is a trivial fixed point at the origin $(r,s)=(0,0)$,
which is of no interest.

\begin{enumerate}
\item[(1)]
With a small rate of back reaction, $\lambda < k_2c_0^2/8$, there are other
six fixed points: 
two on the $r$ axis, U$_1:~(A_-, 0)$ and S$_1:~(A_+, 0)$,
two on the $s$ axis, U$_2:~(0, A_-)$ and S$_2:~(0, A_+)$,
and two symmetric ones, U$_3:~(B_-, B_-)$ and U$_4:~(B_+, B_+)$, 
with 
\begin{align*}
A_{\pm}&=\frac{c_0}{2}\left(1\pm\sqrt{1-\frac{4\lambda}{k_2c_0^2}}\right)\nn
B_{\pm}&=\frac{c_0}{4}\left(1\pm\sqrt{1-\frac{8\lambda}{k_2c_0^2}}\right) .
\end{align*}
As an illustration shown in Fig.5(c),
we use parameter values $k_2c_0^2=1,~\lambda=0.1$, which gives fixed points 
U$_{1-4}$ and S$_{1,2}$ specified by the values,
$A_+=0.8873c_0=c_0-A_-$, and $B_+=0.3618c_0=c_0/2-B_-$.
The stable fixed points S$_1$ and S$_2$ correspond to the states
with complete homochirality with $\phi=\pm 1$.
Except the region around the trivial fixed point O,
the phase flow is attracted to one of these two homochiral states, 
S$_1$ or S$_2$.
Generally, the region of initial points which will eventually be attracted to
a certain fixed point is called the basin of attraction.
In the present case, the $r-s$ phase space is divided into three
basins of attractions to S$_1$, S$_2$, and O, by separatrixes U$_3$U$_1$,
U$_3$U$_2$, and U$_3$M.

The time evolution of the order parameter $\phi$ is depicted in Fig.6(a)
with the parameters $k_0=k_1=0, k_2c_0^2=1, \lambda c_0=0.1$,
the same ones used to obtain the flow diagram in Fig.5(c).
The initial condition is $r_0=0.2+0.001, s_0=0.2$.
The flow is almost symmetric as $r \approx s$ first and 
approaches to the symmetric fixed point U$_4$, but since
U$_4$ is unstable $r$ starts to increase at the cost of $s$ and eventually 
$\phi$ approaches the full value, unity.
\item[(2)]
As the back reaction rate increases, two symmetric fixed points
U$_3$ and U$_4$ merge to disappear for $\lambda \ge k_2 c_0^2/4$.
There remains two  homochiral fixed points S$_1$ and S$_2$. 
Their basins of attraction are close to the $r$ or $s$ axes.
Close to the diagonal $r=s$ the flow runs to the trivial
fixed point O. 
The separatrixes of three basins of attractions are U$_1$M$_1$ and U$_2$M$_2$.
With the parameter values $k_2c_0^2=1,~\lambda=0.2$, 
fixed points S$_1$ and S$_2$ are determined with parameters
$A_+=0.7236c_0$ and $A_-=c_0-A_+$, as shown in Fig.5(d). 
\item[(3)]
Finally, when the back reaction is too strong as $\lambda \ge
k_2c_0^2/4$, all the flow runs to the origin.
\end{enumerate}

\noindent
$\bullet$\,{\bf Linear and quadratic autocatalysis:}

We consider now the effect of a linear autocatalytic process
in addition to the quadratic autocatalysis $k_2$.
As far as $k_1$ is smaller than $\lambda/c_0$, the origin $r=s=0$ is
attractive, and the schematics of the flow in the $r-s$ phase space
is the same with those for $k_1=0$.
When $k_1$ is larger than $\lambda/c_0$,
the origin $r=s=0$ looses the stability, and there are only three fixed points:
two stable ones on $r$ and $s$ axes S$_1$: $(C,0)$ and S$_2$: $(0,C)$ and
a symmetric and unstable one at U: $(D, D)$
with 
\begin{align*}
C&= \frac{k_2c_0-k_1+\sqrt{(k_2c_0+k_1)^2-4k_2 \lambda}}{2k_2}
\\
D&= \frac{k_2c_0-2 k_1+\sqrt{(k_2c_0+2 k_1)^2- 8 k_2 \lambda}}{4k_2} .
\end{align*}
and
the typical flow in this case with 
$k_0=0, k_1c_0 =0.4, k_2c_0^2=1, \lambda=0.3 $ is shown in Fig.5(e), 
which looks 
similar to the case with the cross inhibition
shown in Fig. 3(b): complete homochiral states, $\phi=\pm 1$, 
is attained  asymptotically even though not all the A material is consumed. 

\noindent
$\bullet$\,{\bf Autocatalysis with nonautocatalytic production}

We found that the quadratic autocatalysis with back reaction permits to
attain the homochiral situation, if there are a few C molecules initially
outside 
the basin of attractive to the origin $r=s=0$.
In order to actuate the autocatalysis, however, one has to provide initial
C molecules. Therefore, nonautocatalytic production with a finite $k_0$
seems necessary.
We now consider cases when the random creation of racemic mixtures by $k_0$
 coexist with the autocatalytic production as well as the back reaction.

\begin{enumerate}
\item[(1)]
With a linear autocatalysis $k_0 >0,~k_1>0, k_2=0, \lambda>0$,
the fixed line for $k_0=0$ in Fig.5(b) is replaced by a symmetric fixed point 
at
\begin{align} 
r=s=&\frac{1}{4k_1}\Big\{ k_1c_0-2k_0-\lambda
\nonumber \\
&+\sqrt{(k_1c_0-2k_0-\lambda)^2+8k_0k_1c_0)}\Big\}
\end{align}
which is stable and $\phi=0$.
Random racemization effect by $k_0$ dominates and recovers 
the symmetric state.
\item[(2)]
With a quadratic autocatalysis
$k_0 >0,~k_2>0,~k_1=0$, the situation is rather complex.
Since $f(0)\ne 0$, all the fixed points stay within the triangle (2);
$a,~r,~s \ne 0$.
Then the fixed points satisfy the relation
\begin{align} 
\frac{k_0+k_2r^2}{r}&=\frac{\lambda}{a}=\frac{k_0+k_2s^2}{s}
\nonumber \\
& \rightarrow~~ 
(r-s)(k_0-k_2rs)=0
\end{align}
The symmetry broken states stay on a hyperbola $rs=k_0/k_2$.
They satisfy further the relation
\begin{align} 
\Lambda_a(r) &\equiv k_2 \left[ \frac{c_0^2}{4} -
\left( r-\frac{c_0}{2}+ \frac{k_0}{k_2 r}\right)^2 \right] 
\nonumber \\
& =\lambda >0 .
\label{eq18}
\end{align}
If $k_0$ is smaller than $k_2c_0^2/16$, the 
function $\Lambda_a(r)$ in
 eq.(18)
has two peaks with a height $\lambda_{\mathrm max}=k_2c_0^2/4$,
as shown in Fig.7(a).
It also has a local minimum value 
$\lambda_{\mathrm min}= 2 \sqrt{k_0} (c_0 \sqrt{k_2}-2\sqrt{k_0})$
at $r=\sqrt{k_0/k_2}$.
Then for small $\lambda<\lambda_{\mathrm min}$ there are two solutions 
of asymmetric fixed points
which are stable,
for intermediate case with 
$\lambda_{\mathrm min}< \lambda< \lambda_{\mathrm max}$ there
are four solutions
two among them are stable and two other are unstable,
and for large 
$\lambda > \lambda_{\mathrm max}$ there is none.

If $k_0$ lies between $k_2c_0^2/16$ and $k_2c_0^2/4$, 
$\Lambda_a(r)$ has 
 only a single peak with a value 
$\lambda_{\mathrm max}'= 2 \sqrt{k_0} (c_0 \sqrt{k_2}-2\sqrt{k_0})$.
Therefore the number of solutions are two for small 
$\lambda<\lambda_{\mathrm min}'$ which are stable, 
and none for large $\lambda$.

Further larger $k_0>k_2c_0^2/4$, there is no asymmetric fixed point for
any $\lambda$.
The last result is already drawn by the time evolution of $\phi$ given 
in Eq.(7), which is valid in the present case, too.

\begin{figure}[h]
\begin{center} 
\includegraphics[width=0.45\linewidth]{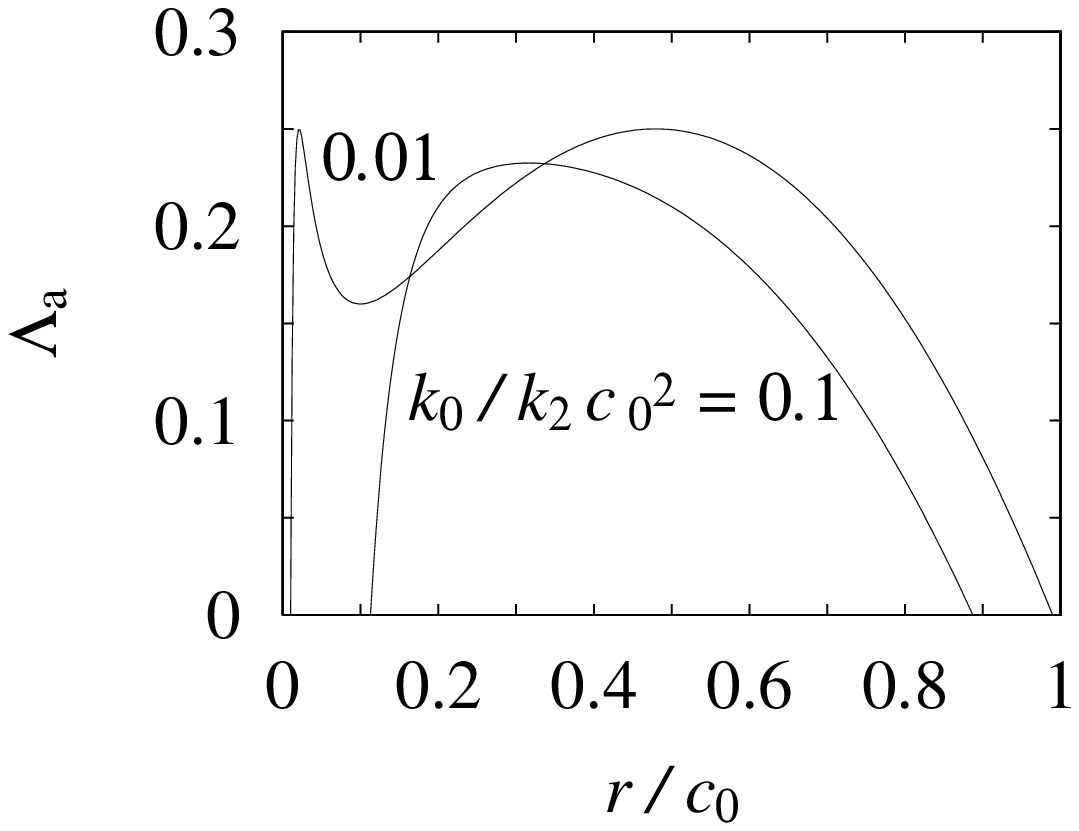}
\includegraphics[width=0.45\linewidth]{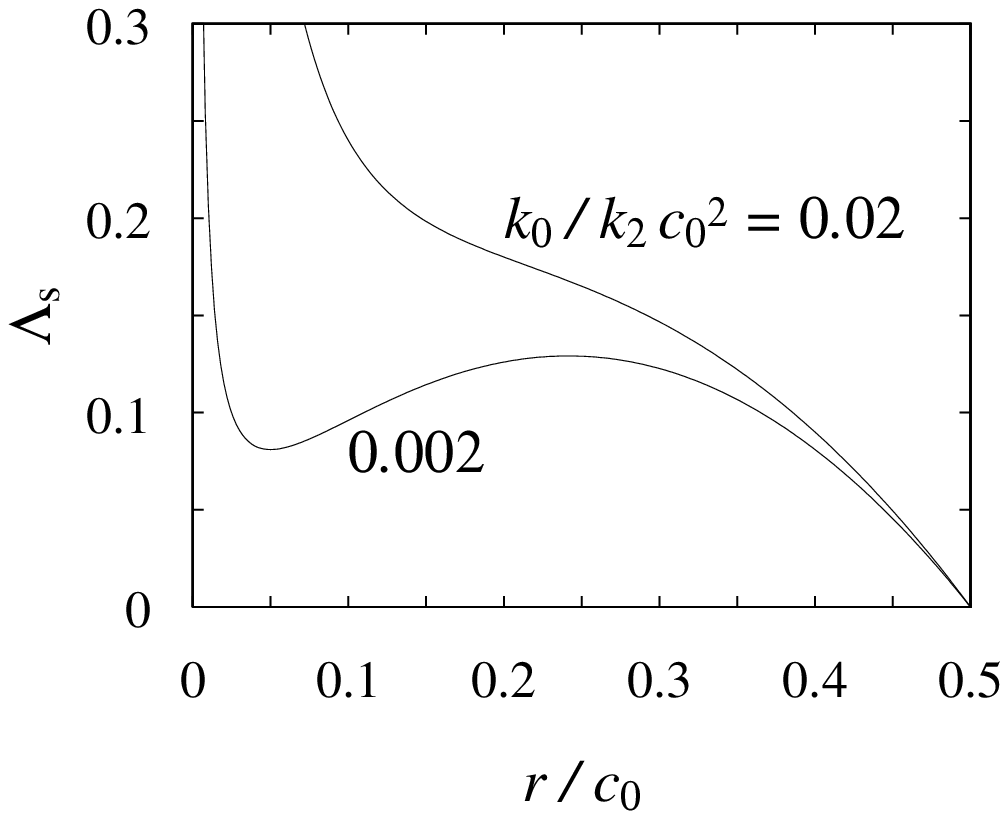}
\end{center} 
\caption{Functions (a) $\Lambda_a(r)$ and (b) $\Lambda_s(r)$ to
determine (a) asymmetric and (b) symmetric fixed points.
}
\label{fig7}
\end{figure}

As for the symmetric solutions with  $r=s$, the effect of random creation $k_0$
is more effective. 
Since the fixed points have to satisfy a relation
\begin{align} 
\Lambda_s(r)  \equiv & \frac{k_2c_0^3}{r} \Big[ -2 \Big(\frac{r}{c_0}-\frac{1}{6} \Big)^3 
\nonumber \\
& + 
\Big( \frac{k_0}{k_2c_0^2}-\frac{1}{108} \Big) \Big] 
+ 2 \Big( \frac{k_2c_0^2}{12}-k_0 \Big) 
\nonumber \\
= & \lambda ,
\end{align}
the number of fixed points reduces to one 
which is unstable
if $k_0$ is larger than
$k_2c_0^2/108$, as shown in Fig.7(b).
For a small rate $k_0 \le k_2c_0^2/108$, the
numbers of symmetric fixed points depends on $\lambda$.
For a small $\lambda$, there is one unstable, for an intermediate value of
$\lambda$ there are three
among which one is stable and two unstable,
 and for a large $\lambda$ there remains only one
stable one.
For example, by adding  $k_0$ as small as $k_0=0.05( > k_2c_0^2/108)$ 
to the situation
shown in Fig.5(c), the origin looses the stability and there remains only 
a single symmetric fixed point which is unstable.
Therefore, the flow diagram reduces to the one shown in Fig.5(f).
The achieved ee is not complete, but is very large if $k_0$ is
small, as shown in Fig.6(b).
There the initial condition is chosen as $r_0=0.1+0.001$ and $s=0.1$,
but the final $\phi$ deos not depend on the initial values of $r_0$ or
$s_0$.
\end{enumerate}

\section{Summary and Discussions}

We studied various autocatalytic models with back reaction or with
cross inhibition processes 
to  search for the conditions to achieve the  chirality selection 
in a closed system. The results are summarized in the 
Table 1. 

\begin{table}[h]
\begin{center}
\begin{tabular}{|c||c|c|c||c|c||c|c|}\hline 
&$k_0$ & $k_1$ & $k_2$ & $\lambda$ & $\mu$ & SB & section  \\ \hline \hline
(1)&$\ge$ & $\ge$ & $\ge$ &     $0$   & $0$   & S  & \S 2   \\ \hline \hline
(2)&$\ge$ & $0$   &  $0$  & $0$   & \textcolor{red}{$>$}& S  & \S 3 \\  \hline
(3)&$\ge$ & \textcolor{red}{$>$}& $\ge$ & $0$ & \textcolor{red}{$>$} &  B& \S 3 \\ \hline \hline
(4)&$\ge$ & $\ge$ & $0$ & \textcolor{red}{$>$} & $0$ & S & \S 4 \\ \hline
(5)&$\ge$ & $\ge$ & \textcolor{red}{$>$} & \textcolor{red}{$>$} & $0$ & B & \S 4 \\ \hline \hline

\multicolumn{8}{|c|}{$\dot{r}=(k_0+k_1r+k_2r^2)a-\lambda r-\mu rs$}\\ 
\multicolumn{8}{|c|}{$\dot{s}=(k_0+k_1s+k_2s^2)a-\lambda s-\mu rs$}\\ 
\multicolumn{8}{|l|}{~\quad$a=c_0-r-s$}\\ \hline
\end{tabular}
\caption{
Classification of reaction models with parameters defined in the bottom column.
Symbol $\ge$ means that the value is non-negative 
and \textcolor{red}{$>$} means that the value must be 
positive.  
B means that the final state is chiral symmetry breaking and 
S means otherwise. 
Details for each entry are explained in the indicated sections.
}
\end{center}
\end{table}

If the production of chiral molecules R and S
from the achiral material A involves a linear autocatalysis,
the cross inhibition between R and S is necessary to realise the homochirality,
as is summarized for the cases (1)-(4) in Table 1.
The case (3) corresponds to an extension of Frank's model to a closed 
chemical system.

With a quadratic autocatalysis, the amplification of enantiomeric excess (ee) 
is achieved without any destruction process, 
but the steady state can have any finite value of ee ($-1\le \phi \le 1$),
as the stable fixed line exists extending from $\phi=1$ to $\phi=-1$.
 In fact the final value of ee depends on the details of the
initial values of chiral as well as achiral materials' concentrations.
Therefore we do not regard this case (1) 
as the chiral symmetry breaking in a restricted sense.

With a linear back reaction process and a quadratic autocatalysis as a special case of (5), the complete homochirality is possible to
realize 
when the rate of back reaction is sufficiently small. 
The minority enantiomer is decomposed 
back to the material A and recycled to produce the majority 
enantiomer and vice versa. 
This decomposition process works more effectively for the minority because of
the suppressed production due to the nonlinear autoctalysis.

However, with only autocatalytic processes, 
the initial production of chiral molecules
is impossible. Therefore, the random production of chiral molecules by
the nonautocatalytic process described by a nonzero value of coefficent $k_0$ is indispensable. For a small value of $k_0$, the phase space structure 
of the flow and fixed points is not qualitatively affected and resembles with
that of vanishing $k_0$ case. However, for $k_0$ greater than the critical value $k_2c_0^2/108$, this process affects the structure of racemic fixed points so drastically 
that the system can have 
only almost homochiral states as stable fixed points, as shown in Fig.5(f).

We have studied two types of simple models (3) and (5) where 
the chiral symmetry can be broken in the closed system 
and complete or almost complete homochiral states can be achieved.
One type of the model with the cross inhibition (3) seems powerful 
for this purpose, but the chemical system which shows this process 
has not been found so far and it is very interesting to find actual chemical system. On the other hand, the nonlinear autocatalytic system is found by Soai's group and thus, there is a real possibility for a nonlinear autocatalytic model with back reaction (5) to exist. 
But we are not certain whether the back reaction proceeds in a measurerable time
for the experiment.

From these analysis of present simple models based on closed systems, we would like to venture to speculate a possible scenario for the initial preparation of constituent materials of life
as follows. In a confined nook and cranny somewhere in the premordial sea,
there was a very favorable conditions for the autocatalytic reactions to take place and after a sufficiently long time, 
the initial homochiral selection was established. Once this selection was completed in this tiny closed corner, the chiral molecules began to spread out and to conquer eventually the surrounding region and futher all over the earth.

So far, we have assumed that the system is homogenious, such that the
whole system is described by only a space-independent concentrations.
In the actual situations, the concentrations should vary in space.
We have introduced a lattice model for the reaction-diffusion system
and studied the spreading of chiral symmetry breaking.
The interested readers are invited to refer
to the original paper.\cite{saito+04b}

\noindent
{\bf Acknowledgements}

The work is supported  by the Gakuji Shinkou Shikin, Keio University.

\renewcommand{\bibname}{References}


\begin{thebibliography}{99}

\bibitem{bonner88}
W. A. Bonner: {\it Topics Stereochem.} {\bf 18}  (1988) 1.


\bibitem{mason+85}
S. F. Mason and G. E. Tranter: {\it Proc. R. Soc. Lond.} {\bf A 397} (1985) 45.

\bibitem{kondepudi+85}
D. K. Kondepudi and G. W. Nelson: {\it Nature} {\bf 314} (1985) 438.

\bibitem{meiring87}
W. J. Meiring: {\it Nature} {\bf 329} (1987) 712.

\bibitem{bada95}
J. L. Bada: {\it Nature} {\bf 374} (1995) 594.

\bibitem{bailey+98}
J. Bailey, A. Chrysostomou, J. H. Hough, T. M. Gledhill, A. McCall, S. Clark, 
F. M\'enard and M. Tamura:
{\it Science} {\bf 281} (1998) 672.

\bibitem{feringa+99}
B. L. Feringa and R. A. van Delden: {\it Angew. Chem. Int. Ed.} {\bf 38} (1999)
3418.

\bibitem{hazen+01}
R. M. Hazen, T. R. Filley and G. A. Goodfriend: {\it Proc. Natl. Acad. Sci.} 
{\bf 98} (2001) 5487.

\bibitem{frank53}
F. C. Frank: {\it Biochimi. Biophys. Acta } {\bf 11}
 (1953) 459.

\bibitem{wynberg89}
H. Wynberg: {\it Chimica} {\bf 43} (1989) 150.

\bibitem{soai+90}
K. Soai, S. Niwa and H. Hori: {\it J. Chem. Soc. Chem. Commun.}  (1990) 982.

\bibitem{soai+95}
K. Soai, T. Shibata, H. Morioka and K. Choji: {\it Nature} {\bf 378} (1995) 767.

\bibitem{sato+01}
I. Sato, D. Omiya, K. Tsukiyama, Y. Ogi and K. Soai: 
{\it Tetrahedron Asymmetry} {\bf 12} (2001) 1965.


\bibitem{sato+03}
I. Sato, D. Omiya, H. Igarashi, K. Kato, Y. Ogi, K. Tsukiyama and K. Soai: 
{\it Tetrahedron Asymmetry} {\bf 14} (2003) 975.

\bibitem{saito+04a}
Y. Saito and H. Hyuga, {\it J. Phys. Soc. Jpn.} {\bf 73} (2004) 33.

\bibitem{saito+04b}
Y. Saito and H. Hyuga, {\it J. Phys. Soc. Jpn.} {\bf 73} (2004) 1685.



\end{thebibliography}
\end{document}